\documentclass[pra,twocolumn,showpacs,nofootinbib,superscriptaddress,floatfix]{revtex4-1}
\usepackage{graphicx}
\usepackage{dcolumn}
\usepackage{multirow}
\usepackage{bm}
\usepackage{comment}
\usepackage{amsmath,amscd,amssymb,color,amsfonts}
\usepackage{epstopdf}
\usepackage{float}
\usepackage{hyperref}
\usepackage{enumerate}

\usepackage{xcolor}
\usepackage{mathtools}
\usepackage{afterpage}
\usepackage{array}
\usepackage[T1]{fontenc}

\newcommand{\RNum}[1]{\uppercase\expandafter{\romannumeral #1\relax}}
\newcolumntype{P}[1]{>{\centering\arraybackslash}p{#1}}
\newcommand\T{\rule{0pt}{2.6ex}}       
\definecolor{darkblue}{rgb}{0,0.0.0,0.4}
\definecolor{darkred}{rgb}{0.5,0,0}
\hypersetup{colorlinks,breaklinks,
linkcolor=darkblue,
urlcolor=darkblue,
anchorcolor=darkblue,
citecolor= darkred,
pdfauthor=JsVgKdAr, pdftitle=JsVgKdAr.ANN}
\begin{document}
\title{Entanglement Classification
of Arbitrary Three-Qubit States via Artificial Neural Networks}
\author{Jorawar Singh}
\email{ph19023@iisermohali.ac.in}
\affiliation{Department of Physical Sciences, Indian
Institute of Science Education \& 
Research Mohali, Sector 81 SAS Nagar, 
Manauli PO 140306 Punjab India.}
\author{Vaishali Gulati}
\email{vaishali@iisermohali.ac.in}
\affiliation{Department of Physical Sciences, Indian
Institute of Science Education \& 
Research Mohali, Sector 81 SAS Nagar, 
Manauli PO 140306 Punjab India.}
\author{Kavita Dorai}
\email{kavita@iisermohali.ac.in}
\affiliation{Department of Physical Sciences, Indian
Institute of Science Education \& 
Research Mohali, Sector 81 SAS Nagar, 
Manauli PO 140306 Punjab India.}
\author{Arvind}
\email{arvind@iisermohali.ac.in}
\affiliation{Department of Physical Sciences, Indian
Institute of Science Education \& 
Research Mohali, Sector 81 SAS Nagar, 
Manauli PO 140306 Punjab India.}
\begin{abstract}
We design and successfully implement artificial neural networks (ANNs) to
detect and classify entanglement for three-qubit systems using  limited
state features.  The overall design principle is a feed forward neural
network (FFNN), with the output layer consisting of a single neuron for
the detection of genuine multipartite entanglement (GME) and six
neurons for the classification problem corresponding to six
entanglement classes under stochastic local operations and classical
communication (SLOCC).  The models are trained and validated on a
simulated dataset of randomly generated states.  We achieve high
accuracy, around 98\%, for detecting GME as well as for SLOCC
classification.  Remarkably, we find that feeding only 7 diagonal
elements of the density matrix into the ANN results in an accuracy
greater than 94\% for both the tasks, showcasing the strength of the
method in reducing the required input data while maintaining efficient
performance.  Reducing the feature set makes it easier to apply ANN
models for entanglement classification, particularly in
resource-constrained environments, without sacrificing accuracy. The
performance of the ANN models was further evaluated by introducing
white noise into the data set, and the results indicate that the models
are robust and are able to well tolerate noise.
\end{abstract}
\maketitle
\section{Introduction}
\label{sec1}
Entanglement, alongwith with other quantum correlations,
represents a fundamental aspect of quantum mechanics that
underpins the superiority of quantum information processing
over classical approaches~\cite{chuang-book}. Entanglement
is an essential ingredient for quantum algorithms such as
Shor’s algorithm and other algorithms which provide an
exponential speedup over their classical
counterparts~\cite{shor1994algorithms,blekos2024review}.
Furthermore, entanglement has been instrumental in enabling a
range of quantum technologies, including quantum
communication\cite{bennett1993teleporting,ekert1991quantum},
super-dense coding\cite{bennett1992communication}, and
quantum metrology\cite{giovannetti2004quantum}. Despite
these advantages, the detection, estimation, and
classification of quantum entanglement can be
computationally formidable~\cite{gharibian2010strong}, and
pose a significant challenge in the
field~\cite{guhne2009entanglement}.

While entanglement detection and classification are
well-characterized for bipartite pure states and for general
states of two-qubit systems due to the availability of
positive partial transpose (PPT)
criteria~\cite{horodecki1996seprability}, they remain open
problems for multipartite  systems as well as for mixed
states of larger-dimensional bipartite systems.  For
instance, all the mixed entangled states for two qutrits are
not known, while  for a 4-qubit system, the total number of
distinct entanglement classes is yet to be fully
delineated~\cite{verstraete2002four,li2007classification}.
Traditional approaches for entanglement detection  utilize
one-way (sufficient) conditions based
on Bell inequalities, concurrence, negativity and
entanglement witnesses.  Oftentimes, these detection schemes
require full state tomography, which is computationally hard
and requires a lot of
resources~\cite{bell1964on,wooters2001entanglement,dong2024quantifying}.

In the present era where large scale data manipulations are
possible, artificial intelligence (AI) provides
innovative approaches to enhance and at times even replace,
existing techniques for computation and decision making
problems.  AI has been successfully utilized across a wide
array of fields, ranging from healthcare
diagnostics~\cite{esteva2019guide} and climate
modeling~\cite{reichstein2019deep} to
chatbots~\cite{brown2020language} and
gaming~\cite{togelius2011search,jordan2015machine}. In the
context of quantum computation and information, 
machine learning (ML) has been
applied to several information processing tasks, including
quantum error correction~\cite{varsamopoulos2018decoding},
optimization~\cite{cerezo2021variational}, quantum
simulation~\cite{Kassal2011annual}, data
compression~\cite{aaronson2004improved}, and verifying
quantum supremacy~\cite{harrow2017quantum}.

ML-based strategies for entanglement detection include artificial neural
network (ANN) models using Bell inequalities for classifying entangled
states~\cite{ma2018transforming}, deep ANN for identifying
entanglement~\cite{mateusz2024data,urena2024entanglement}, and validating the
entanglement class of three-qubit pure states via ANN
models~\cite{gulati2024ann}.  ML has also been used to parameterize quantum
states as neural networks~\cite{harney2020entanglement,harney2021mixed}, to
optimize measurement sets for entangled states~\cite{yosefpor2020finding}, and
to generate entanglement witnesses using support vector
machines~\cite{sanavio2023entanglement,greenwood2023machine}, and using
generative adversarial networks~\cite{chen2020detecting}.

SLOCC (Stochastic Local Operations and Classical
Communication) classification is a framework used to
categorize quantum states based on their equivalence under
local operations combined with classical communication.  Two
states \( |\psi\rangle \) and \( |\phi\rangle \) are
considered equivalent under SLOCC if there is a non-zero
probability of successfully converting \( |\psi\rangle \)
into \( |\phi\rangle \), as well as converting \(
|\phi\rangle \) back into \( |\psi\rangle \) under LOCC
\cite{bennett-pra-2000}. For the three-qubit system that we
study in this paper,
the general form of pure states upto LOCC
can be written in a convenient canonical form, from
which arbitrary states can be generated by the application
of local unitaries~\cite{acin-canonical}:
\begin{eqnarray}
&\!\!\!|\psi\rangle= \lambda_0|000\rangle+\lambda_1 e^{i \varphi}|100\rangle
+\lambda_2|101\rangle+\lambda_3|110\rangle+\lambda_4|111\rangle
\label{eq_acin} \!\!\!\!\!\nonumber \\
&{\displaystyle \sum_{i=0}^{i=4}}|\lambda_{i}|^{2} = 1,
\quad \lambda_{i}
\in \mathbb{R}, \quad \lambda_{i} \geq 0,\quad 0 \leq \varphi
\leq \pi.
\end{eqnarray}
In a three-qubit system, there are six inequivalent SLOCC
classes: Fully separable, where all qubits are separable;
three biseparable classes (BS1, BS2, BS3), where specific
pairs of qubits are entangled while the remaining qubit is
not entangled with them
; and two genuine multipartite classes, the GHZ and
W classes, with each class representing different forms of maximal
entanglement~\cite{dur-pra-2000}.

In this study, ANN models were designed and implemented with
two main objectives: ({\bf 1.}) To detect the presence of genuine
multipartite entanglement (GME) in pure three-qubit states,
and ({\bf 2.}) to classify  entanglement in these systems by
distinguishing between six SLOCC classes, namely: fully
separable (SEP), three types of biseparable states (BS1,
BS2, BS3), and maximally entangled W and GHZ states.  We
employ different ANN models for each task. The first ANN
model utilizes all 35 relevant density matrix elements,
while the second model is restricted to the 7 diagonal
elements. The models were trained and tested using
numerically generated arbitrary pure quantum states.  Our
results demonstrate that even with a reduced set of
features, specifically the diagonal elements alone, the ANN
model can still effectively classify complex quantum states,
highlighting the robustness and efficiency of this approach.
It is worth noting that the high efficiency of detection of
GME and classification of entanglement from only the 7
diagonal elements of the density matrix is rather surprising
and points to certain hidden patterns in the data which the
ANN model is able to recognize.

The rest of this paper is structured as follows:
Section~\ref{ann} covers the fundamental concepts necessary
for designing our ANN models, including dataset
generation~\ref{generation}, feature
selection~\ref{feature-selection}, and basics of the ANN
architecture and evaluation~\ref{ann-scheme}.
Section~\ref{results} contains details of our models and
presents the results obtained from the
GME~\ref{classify_gme} and SLOCC~\ref{classify_slocc}
models.  Section~\ref{concl} contains a few conclusions and
future directions.
\section{Essentials of ANN Model Building}
\label{ann}
\subsection{Dataset Generation and Data encoding}
\label{generation}
Random quantum states of the three-qubit system required to train our ANN model
are generated by utilizing the qutip package~\cite{qutip}.  The dataset
aught to encompass various SLOCC 
classes, each corresponding to different types of entanglement
in three-qubit systems, as detailed below.

\vspace*{6pt}
\noindent{\bf Separable States}: These are generated by
taking the tensor product of three independent single-qubit states.
These states are unentangled across all qubits
and provide examples of fully separable quantum states in
our training data.

\vspace*{6pt}
\noindent{\bf Biseparable States}: These states are
partially entangled, with only two of the three qubits
entangled. We further categorize them into three subsets based on
which qubits are entangled:

\vspace*{3pt}
\noindent Biseparable 1 (BS1): States in this set are
generated as a tensor product of a 2-dimensional ket with a
4-dimensional ket (2$\otimes$4), representing entanglement
between the second and third qubits.

\vspace*{3pt}
\noindent Biseparable 3 (BS3): Here, we generate states as
4$\otimes$2 kets, where the first two qubits are entangled.

\vspace*{3pt}
\noindent Biseparable 2 (BS2): For this set, we start by
creating 4$\otimes$2 states in a ACB qubit order, then apply
a swap operation between qubits B and C to re-order qubits to
the standard ABC configuration, where qubits A and C are
entangled.

\vspace*{6pt}
\noindent {\bf GHZ States}: GHZ states are maximally
entangled across all three qubits, where no two qubits
are individually entangled. These states are represented
by random kets of dimension 8.

\vspace*{6pt}
\noindent {\bf W States}: Although W states are also
maximally entangled across all three qubits and belong to
the 8-dimensional Hilbert space, they form a set of measure
zero~\cite{Enriquez-entropy-2018}, meaning random sampling
of 8-dimensional kets will not produce W states. To ensure
the representation of this set, we start with the standard W
state and apply local unitary transformations, which
maintain W-state entanglement properties while generating
variations.

To ensure that the ANN model is exposed to a wide variety of
state configurations, we apply local unitary transformations
to the randomly generated states from each SLOCC class
(separable, biseparable, and GHZ). In a three-qubit setup,
there are 8 possible ways to apply local unitary
transformations to a given state:
$\{\mathbb{I}\otimes\mathbb{I}\otimes\mathbb{I},
U_i\otimes\mathbb{I}\otimes\mathbb{I},..., U_i\otimes
U_j\otimes U_k\}$, with each $U$ generated randomly. For
each SLOCC class, we generate 125,000 states using each of
the 8 local unitary transformations, resulting in a
total of 1,000,000 states. The base state for each
transformation is generated as described earlier.  This
augmentation step enhances the diversity and
generalizability of the dataset, providing the ANN model
with a wide range of instances from each SLOCC class,
thereby improving its learning performance. For both
tasks-SLOCC classification and GME detection, 
the same 1,000,000 states are used for
training.

Having generated the input-output pairs for the ANN
training, we need to encode them in a way that the network
understands: The $8\times8$ complex valued density matrix
$\rho$ is split into two $8\times8$ real valued matrices
comprising of the real and imaginary components of $\rho$.
These $8\times8$ matrices are then flattened into column
vectors, resulting in $\vec{V}_\mathbb{R}(\rho)$ and
$\vec{V}_\mathbb{I}(\rho)$ of size $64\times1$.  These
vectors are further combined into a single column vector of
size $128\times1$: $\vec{F} =
[\vec{V}_{\mathbb{R}}(\rho)~~\vec{V_{\mathbb{I}}}(\rho)]$
which becomes the input feature vector to the ANN model.

For the GME/Non-GME identification problem, the state labels are encoded as
binary numbers in a single-element vector $\vec{L}(\rho)$,
with 1 indicating GME and 0 indicating Non-GME.  For the
SLOCC classification problem, the vector $\vec{L}(\rho)$ is
a six-element column vector, with each position
corresponding to a SLOCC label: [SEP, BS1, BS2, BS3, W,
GHZ]. Using one-hot encoding ($\vec{L}(\rho)$ =
[$L_{i}(\rho)$], $L_{i}(\rho)$ $\in  \left \{ 0,1\right \}$
and $\sum_{i=1}^{6}L_{i}(\rho)=1$), we represent the class
labels, assigning a value of 1 or 0 at the position
corresponding to the SLOCC label.  This vector
$[\vec{L}(\rho)]$ becomes the output for a given input
$\vec{F}$.

\begin{figure*}
\centering
\includegraphics[scale=1]{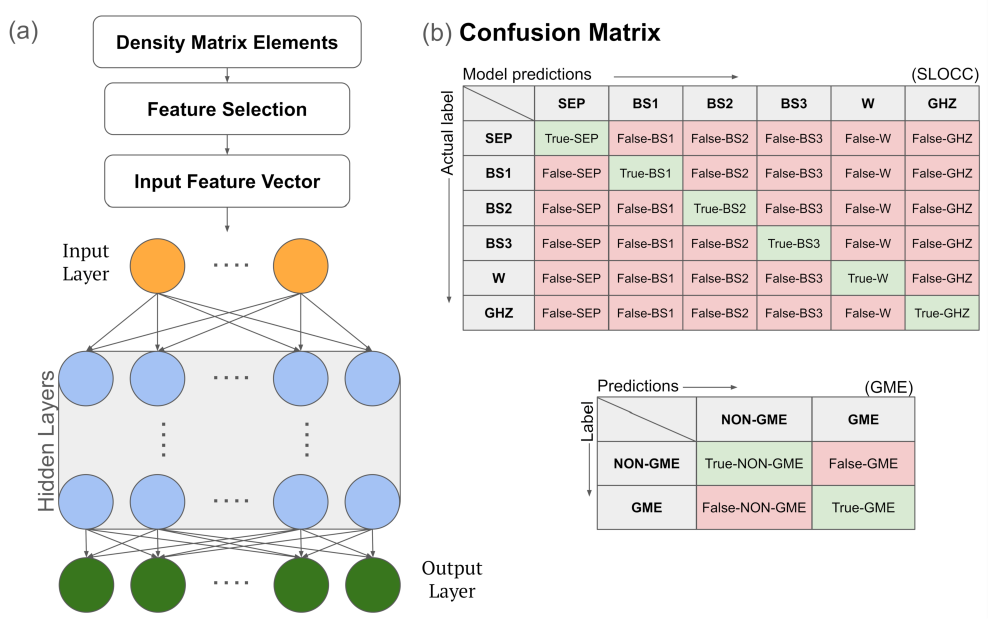}
\caption{(Color online) The schematic of 
the ANN-based entanglement classification protocol is
shown in panel (a). For both GME and SLOCC tasks, the raw
complex density matrix data is first simplified to real numbers,
followed by feature selection to remove redundant and irrelevant
features. The selected features, based on univariate selection, are
then fed into the respective GME and SLOCC ANN models.  The input layer
for the models is shown in orange, the hidden layers are in blue, and
the output layer is in green. Panel (b) depicts the confusion matrices
for the two models, which can be used to access the overall
performance of the model as well 
as its effectiveness for each individual target
class.}
\label{ANN-figure} 
\end{figure*}

\subsection{Feature Selection}
\label{feature-selection}
Feature selection refers to the removal of redundant and irrelevant features
from the dataset, to improve the training, performance, and efficiency of the
ANN model.  
The input data to our ANN model are elements
of a density matrix, corresponding to a three-qubit quantum state.  
Given that the density matrix is
Hermitian with a trace of 1, we discard the upper triangular part, along with
one element from the principal diagonal, without any loss of information. This
reduces the matrix to 28 complex elements and 7 real elements, which translates
into 63 real numbers used as input features for the ANN.  We further apply
univariate selection over these remaining 63 features to assign importance
scores to each of the features.

Univariate selection is a feature selection method that evaluates the
relationship between an individual feature and the target variable, and applies
statistical tests to estimate the importance of each feature for the
classification task at hand. We use the ANOVA (Analysis of Variance) F-test to
assign importance scores to our
features~\cite{betty-epm-1974,gelman-as-2005,ostertagova-ajme-2013}.  The ANOVA
F-test partitions the dataset according to class labels and computes the mean
value of each feature within these groups. It then compares two sources of
variability: between-group variance (the extent to which the mean values differ
across the groups) and within-group variance (the variability of feature values
within each group). A feature is considered significant if the between-group
variance is substantially larger than the within-group variance, indicating
that the feature effectively distinguishes between classes.

\subsection{Model Construction and Evaluation}
\label{ann-scheme}
A basic feed-forward neural network (FFNN) consists of three main components:
an input layer, responsible for loading the features; one or more hidden
layers, where the network learns relationships through the adjustment of
weights and biases; and an output layer, which predicts the labels
corresponding to the input.  Each layer applies an activation function (such as
ReLU or sigmoid) which introduces non-linearity to the network, enabling it to
model complex patterns in the data~\cite{bern-cam-2022}.

To train the FFNN, we split the dataset into a 80\% training
set and 20\% validation set. While the model uses states in
the training set to learn the patterns in the data, the
validation set states are used to observe the learning
process and ensure that the model does not overfit the data.
The FFNN is trained using a process called backpropagation,
which adjusts the model’s weights and biases to improve
performance. Backpropagation operates by calculating a loss
function (or cost function) that measures the error between
the predicted and true labels for the training dataset.
This loss is minimized by iteratively updating the weights
of the network over a series of passes
through the training data, known as
epochs~\cite{bern-cam-2022}. As training progresses, the
model refines its predictions by learning from errors,
aiming to reduce the overall loss.

The training process is monitored by observing the trends
in the values of the indicators: training loss, training accuracy,
validation loss, and validation accuracy with respect to the
epochs (total training time). These indicators are used to
tune hyperparameters such as the number of neurons, number
of hidden layers, learning rate, number of epochs, and batch size.
The values of these indicators reflect the optimization process, which
is performed using the stochastic gradient descent technique
called ``Adam'' with a fixed learning
rate\cite{sun-ieee-2020}. The goal of the optimization is to
minimize the loss values over the training set.
While a proper training of the model is reflected
by a simultaneous decrease in training and validation loss,
overfitting typically manifests as increasing validation
loss coupled with decreasing training loss, signaling that
the model is learning the specific details of the training
data (including noise) rather than general patterns.
 Overfitting can generally be avoided by not using too
complex a model and limiting the number of epochs.

The performance of a trained model is commonly assessed
using the accuracy metric ($A$). Accuracy measures how well
the model’s predictions align with the true labels by
calculating the ratio of correct predictions to the total
number of predictions made. It is expressed as:
\begin{equation}
A = \frac{\sum \text{Correct predictions per class}}{\text{Total predictions}}
\label{eq-metric-accuracy}
\end{equation}

In this equation, the numerator sums the number of correct predictions for each
class, and the denominator represents the total number of predictions,
including both correct and incorrect ones.

Figure~\ref{ANN-figure} presents an overview of the data
processing pipeline and the structure of the ANN models. As
described in previous subsections, feature selection is
applied to eliminate redundant features from the input
density matrix, $\rho$, and to rank the remaining features
in descending order of their importance, which is assessed
using univariate selection via the ANOVA F-test. This
process ensures that only relevant features are retained,
reducing the dimensionality of the input space and
simplifying the removal of irrelevant features, if any. The
selected features are subsequently fed into the ANN model to
perform the classification task.

Figure~\ref{ANN-figure}(a) illustrates the general ANN
architecture for the entanglement classification problem.
The architecture consists of an input layer with $N$
neurons, where $N$ corresponds to the number of features
extracted from the density matrix, followed by a series of
hidden layers with a varying number of neurons, and finally
the output layer with number of neurons depending upon the
classification problem. There are 6 neurons in the output
layer for SLOCC classification and 1 in case of GME
detection.

The number of hidden layers and neurons in the hidden layers
is initially set using a heuristic, where, 
the number of neurons in the $i$-th hidden
layer, $N^i$, is set to approximately half of the number
of neurons in the previous layer, $N^{i-1}$, following the formula:
\[
N^{i} = \frac{(N^{i-1} + N^{i-1}\text{mod}~ 2)}{2}
\]
where $N^0(=N)$ represents the number of neurons in the input layer. This
configuration is then fine-tuned based on the complexity of the problem, as
outlined in \cite{gaurang-ijcte-2011}.  The heuristic aims to gradually reduce
the number of neurons in deeper layers, balancing the complexity of the model
while preventing overfitting. The formula halves the number of neurons with a
small adjustment for even or odd numbers (due to the $\mod$ operation). This
setup provides a reasonable starting point, which is later refined through
trial and error depending on the specific problem.

The model classifies input density matrices, $\rho$, as
either exhibiting or not exhibiting GME, with the final
output layer consisting of a single neuron that outputs a
probability, $0 \leq p \leq 1$.  This probability indicates
the likelihood that the input state belongs to the GME
class. A probability value closer to 1 suggests the state
has GME, while a value closer to 0 indicates that it does
not.

For the SLOCC entanglement classification problem, the
general architecture remains the same, with input layer of
$N$ neurons, followed by a series of hidden layers.  The
output layer for this task consists of 6 neurons, each
representing one of the six SLOCC entanglement classes: SEP,
BS1, BS2, BS3, W, and GHZ. The output of the ANN is a
probability distribution ${p_1, p_2, p_3, p_4, p_5, p_6}$,
where each $p_i$ denotes the probability that the input
state belongs to class $i$.  The class with the highest
probability is selected as the predicted label. For
instance, if $p_4$ is the largest, the input state is
classified as belonging to the BS3 class.
In both models, the predicted label is then compared to the
true class label of the state. If the predicted label
matches the true label, the state is considered correctly
classified, and the accuracy for that state is recorded as
1. If the predicted label differs from the true label, the
state is misclassified, and the accuracy is 0.

This classification performance, including the 
errors, of the model for each output class is visualized using a
confusion matrix, as shown in Figure~\ref{ANN-figure}(b), which 
compares the predicted and true labels of the states
classified by the ANN. Each entry in the matrix represents
the number of states for which the predicted class label
(assigned by the ANN) matches or differs from the true class
label. The confusion matrix also allows an easy calculation
of the accuracy $A$ of the model - by summing the main
diagonal entries and dividing it by the total number of
entries~\cite{diego-ai-2022,bern-cam-2022}.  The smaller
table in Figure~\ref{ANN-figure}(b) shows the confusion
matrix for GME classification.  Misclassifications are
categorized as `False Non-GME'—instances where GME cases
were incorrectly predicted as non-GME—and `False GME'—cases
where non-GME instances were mistakenly classified as GME.
For the SLOCC case, (SLOCC Table in Figure
\ref{ANN-figure}(b)), a correctly classified state is
indicated by entries like "True-SEP", meaning the true label
and the predicted label are both SEP (separable).
Similarly, "True-BS1", "True-BS2", etc., represent correct
classifications for other SLOCC classes. Misclassified
states are denoted by entries such as "False-SEP",
indicating that a state with a true label other than SEP
state was misclassified as SEP by the ANN.  For each SLOCC
entanglement class, the confusion matrix tracks both
correctly and incorrectly predicted labels. This allows us
to evaluate the performance of the ANN across all classes
and identify which entanglement categories are more prone to
classification errors.
\section{Results and Discussion} 
\label{results}
\subsection{Reduction of Selected Features}
\label{univariate_7}
To address the classification problem, we numerically
generated a dataset of 6,000,000 quantum states, with each
class represented by 1,000,000 states. To evaluate the
relative importance of each feature for classification, we
employed the ``f classif'' function along with the ``Select
K Best'' method from the Sklearn library
\cite{scikit-learn}. These methods calculate ANOVA F-values,
which provide scores indicating each feature's significance
in the classification task. Table~\ref{table_score_1}
presents a sample of the top-ranking features and their
corresponding ANOVA F-values from one specific run, out of
many performed. After performing multiple runs with
different datasets, we observed a consistent pattern: the
top 7 features always emerged from the specific density
matrix elements—$\mathbf{Re}(\rho_{00})$,
$\mathbf{Re}(\rho_{11})$, $\mathbf{Re}(\rho_{22})$,
$\mathbf{Re}(\rho_{33})$, $\mathbf{Re}(\rho_{44})$,
$\mathbf{Re}(\rho_{55})$, and $\mathbf{Re}(\rho_{66})$.
While the ranking of these features varied slightly between
runs, these 7 features consistently occupied the top
positions. Additionally, there was a marked drop in the
ANOVA F-values after the 7th feature, indicating a
significant difference in importance between these top
features and the remaining 56. We then evaluate whether an
ANN model using only these top 7 features can achieve
performance comparable to a model that uses all 63 features
for both classification problems.

\begin{table}
\begin{ruledtabular}
\begin{tabular}{ccrr}
Feature No. & Feature & SLOCC & GME      \\
\hline \T
1 & $ \mathbf{Re}(\rho_{22}) $ & 20275.73 & 39573.50 \\
2 & $ \mathbf{Re}(\rho_{44}) $ & 20192.56 & 39899.69 \\
3 & $ \mathbf{Re}(\rho_{11}) $ & 20120.34 & 39422.10 \\
4 & $ \mathbf{Re}(\rho_{55}) $ & 6768.37 & 13527.75 \\
5 & $ \mathbf{Re}(\rho_{66}) $ & 6724.48 & 13335.46 \\
6 & $ \mathbf{Re}(\rho_{33}) $ & 6687.62 & 13151.14 \\
7 & $ \mathbf{Re}(\rho_{00}) $ & 2288.79 & 4650.39 \\
8 & $ \mathbf{Im}(\rho_{07}) $ & 2.75 & 0.13 \\
9 & $ \mathbf{Re}(\rho_{23}) $ & 2.45 & 0.06 \\
10 & $ \mathbf{Im}(\rho_{24}) $ & 2.14 & 1.73 \\
\end{tabular}
\end{ruledtabular}
\caption{ANOVA F-values (Feature Scores) for top 10 features
for the SLOCC and GME ANN models.}
\label{table_score_1}
\end{table}

\subsection{GME Classification via ANN}
\label{classify_gme}
We use the Keras API~\cite{keras} with 
Tensorflow backend~\cite{abadi-tf-2015} for the construction,
optimization, and analysis of the ANN models.
We begin by constructing an artificial neural network (ANN)
model utilizing all $N = 63$ input features. After extensive
hyperparameter tuning to optimize accuracy, the resulting
model architecture consists of layers with the following
neuron structure: $N-64-32-16-8-4-O$, where the output layer
has a single neuron ($O = 1$). Each hidden layer employs the
ReLU activation function $(ReLU(x)=max(0,x))$, while the
output layer uses the sigmoid activation function, defined
as:
\begin{equation}
    \sigma(x) = \frac{1}{1 + e^{-x}}
    \label{eq_sigmoid}
\end{equation}
In this equation, $x$ represents the weighted average of the
inputs passed from the final hidden layer, and $\sigma(x)$
outputs the probability that the given input belongs to the
target class. This choice of activation function in the
output layer is standard for binary classification tasks,
where the output is interpreted as a probability.

The model's loss function is binary cross-entropy, which is
commonly used for binary classification problems. It
quantifies the difference between the predicted
probabilities and the actual class labels. The binary
cross-entropy loss is given by:
\begin{equation}
\text{Loss} = -\frac{1}{N}\sum_{i=1}^{N} \left[ y_i \ln(p_i)
+ (1 - y_i) \ln(1 - p_i) \right]
\end{equation}
where $N$ is the total number of data points (or states),
$y_i = L_i(\rho)$ is the actual class label for the $i$-th
data point, and $p_i$ is the predicted probability of that
point belonging to the target class, i.e. $\sigma(x)$.

The model is trained using the Adam optimizer, with a
learning rate of 0.0025 and a batch size of 5000. Training
is conducted over 100 epochs, leading to an average accuracy
of $0.998 \pm 0.001$ on the test set, indicating highly
effective classification performance.

Based on the results from univariate feature selection with
ANOVA, we assessed the model's performance using only the
top 7 features, specifically the diagonal elements of the
density matrix. Using these 7 features with the original
model architecture and hyperparameters, we observed a
decrease in accuracy of approximately 2\%. However, after
fine-tuning the model to better accommodate the reduced
input dimension (now structured as $N-64-32-16-O$) and
extending the training duration to 200 epochs, the accuracy
loss was nearly eliminated. This resulted in an average test
set accuracy of $0.992 \pm 0.002$, indicating that the model
performs well even with a reduced set of input features.

\begin{figure}[h!]
\includegraphics[scale=1.3]{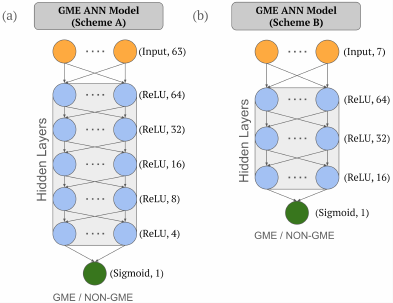}
\caption{GME Classification ANN Schemes: Scheme A  (shown in panel (a))
utilizes 63 input features, while Scheme B (shown in panel (b)) uses 7
input features.  Both schemes predict the probability of the input
state having genuine multipartite entanglement (GME) or not. The input
layer is represented with orange circles, the hidden layers with blue,
and the output layer with green.}
\label{GME-ANN-figure} 
\end{figure}

The ANN architecture for the GME classification problem is
illustrated in Scheme A and Scheme B, which correspond to
models using 63 features and 7 features, respectively, as
shown in Figure \ref{GME-ANN-figure}. 
In these diagrams, orange circles represent the input layer,
while the hidden layers are depicted in blue, and the output
layer is shown in green. 
Each layer is labeled with its activation function and the
number of neurons it contains, which are specified in
brackets.

Table \ref{table_averages} presents the average performance of ANN models in
classifying data into Non-GME and GME categories, using two different feature
sets: 63 features and 7 features. The models demonstrate high classification
accuracy across both categories. For Non-GME data, the model achieves an
accuracy of 0.98 with 63 features and 0.99 with 7 features. For GME data, the
model attains an accuracy of 0.99 with 63 features and 0.98 with 7 features.
The overall average accuracy for the models is 0.99, regardless of whether 63
or 7 features are used. This indicates that the ANN models perform consistently
well in distinguishing between Non-GME and GME states, even when using a
reduced number of features.

\begin{table}[h!]
\begin{tabular}{|c|ccc|c|cc|}
\hline
\multirow{2}{*}{\textbf{Class}} & \multicolumn{3}{c|}{\textbf{SLOCC}} & \multirow{2}{*}{\textbf{Class}} & \multicolumn{2}{P{1.8cm}|}{\textbf{GME}} \T\\ \cline{2-4} \cline{6-7} 
 & \multicolumn{1}{P{1cm}|}{\textbf{63 F}} & \multicolumn{1}{P{1cm}|}{\textbf{7 F}} & \textbf{7 F(Hier.)} &  & \multicolumn{1}{P{0.9cm}|}{\textbf{63 F}} & \textbf{7 F} \T\\ \hline
\textbf{FS} & \multicolumn{1}{c|}{0.96} & \multicolumn{1}{c|}{0.91} & 0.93 & \multirow{4}{*}{\textbf{\begin{tabular}[c]{@{}c@{}}Non\\ GME\end{tabular}}} & \multicolumn{1}{c|}{\multirow{4}{*}{0.98}} & \multirow{4}{*}{0.99} \T\\ \cline{1-4}
\textbf{BS1} & \multicolumn{1}{c|}{0.98} & \multicolumn{1}{c|}{0.95} & 0.97 &  & \multicolumn{1}{c|}{} &  \T\\ \cline{1-4}
\textbf{BS2} & \multicolumn{1}{c|}{0.98} & \multicolumn{1}{c|}{0.96} & 0.96 &  & \multicolumn{1}{c|}{} &  \T\\ \cline{1-4}
\textbf{BS3} & \multicolumn{1}{c|}{0.98} & \multicolumn{1}{c|}{0.96} & 0.96 &  & \multicolumn{1}{c|}{} &  \T\\ \hline
\textbf{W} & \multicolumn{1}{c|}{0.97} & \multicolumn{1}{c|}{0.84} & 0.91 & \multirow{2}{*}{\textbf{GME}} & \multicolumn{1}{c|}{\multirow{2}{*}{0.99}} & \multirow{2}{*}{0.98} \T\\ \cline{1-4}
\textbf{GHZ} & \multicolumn{1}{c|}{0.97} & \multicolumn{1}{c|}{0.86} & 0.93 &  & \multicolumn{1}{c|}{} &  \T\\ \hline
\textbf{Average} & \multicolumn{1}{c|}{0.98} & \multicolumn{1}{c|}{0.91} & 0.94 & \textbf{} & \multicolumn{1}{c|}{0.99} & 0.99 \T\\ \hline
\end{tabular}
\caption{Average performance of SLOCC and GME ANN models across respective classification classes. "63F" and "7F" represent models that use 63 and 7 features, respectively. "7F (hier.)" refers to the hierarchical 3-model SLOCC ANN setup, highlighting the difference in model performance with and without the hierarchical structure.}
\label{table_averages}
\end{table}

\subsection{SLOCC Classification via ANN}
\label{classify_slocc}

Similar to the GME classification case, we start with an ANN model that uses 63 input features. The architecture of the hidden layers is the same as before, comprising a sequence of layers with $64-32-16-8-4$ neurons, each using the ReLU activation function. The output layer consists of 6 neurons and utilizes the softmax activation function, defined as:
\begin{equation}
\sigma(x_i) = \frac{e^{x_i}}{\sum_{j=1}^C e^{x_j}}, \quad \text{for} \quad i = 1, 2, \ldots, C \label{eq_softmax} \end{equation}
Again, $x_{i}$ represents the weighted average of the inputs from the final hidden layer
and $\sigma(x_{i})$  represents the probability of the state belonging to the $i$-th class, with $C$ denoting the total number of classes.

Following the standard choice, we use "categorical cross-entropy" as the loss function, defined as:
\begin{equation} \text{Loss} = -\frac{1}{N}\sum_{j=1}^{N} \left(\sum_{i=1}^{C} y_{i}^{j} \ln(p_i^j)\right) \end{equation}
where 
$N$ is the total number of data points, 
$C$ is the total number of classes,
$y_{i}^j = L_i^j(\rho)$ and $p_i^j = \sigma(x_i)$ are the true class label and predicted probability for the $j$-th data point in the $i$-th class respectively.

The model is trained for 1000 epochs using the Adam optimizer with a learning rate of 0.0025 and a batch size of 5000, resulting in an average test set accuracy of $0.984 \pm 0.002$.

For the SLOCC classification model using only 7 features, the overall accuracy decreased to approximately 86\%. Despite improvements up to 91\% with further hyperparameter tuning, the model's performance for the W and GHZ classes remained around 85\%. The detailed accuracy for each SLOCC class along with the average accuracy of the model is tabulated in Table\ref{table_averages}.

To address this issue, we introduced a hierarchical multi-ANN structure and trained three distinct models:
\begin{enumerate}
    \item \textbf{FS-NOT}: This model has an architecture of
\( N - 32r - 32r - 16l - 16l - 16l - O \), where \( r \) and
\( l \) denote ReLU and linear activations, respectively,
and \( O = 2 \). It classifies inputs as either separable or
not separable.
    \item \textbf{BS-GME}: This model follows the structure
\( N - 32r - 16r - 8r - 4l - O \), with \( O = 4 \). It
classifies inputs into one of the  biseparable classes (BS1,
BS2, BS3) or as GME, which encompasses the W and GHZ
classes. 
    \item \textbf{GHZ-W}: This model features an
architecture of \( N - 128r - 64r - 32r - 16r - 8r - 4l - O
\), where \( O = 2 \). It differentiates between GHZ and W
classes.
\end{enumerate}

\begin{figure}[h!]
\includegraphics[scale=0.95]{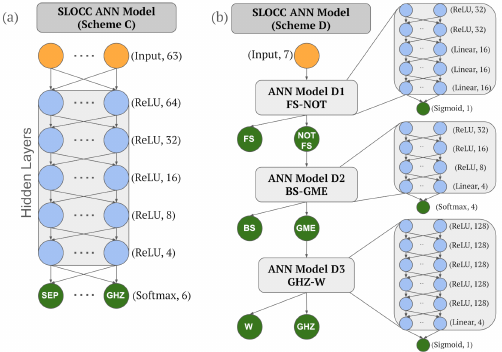}
\caption{SLOCC Classification ANN Schemes: Scheme C (shown in panel (a))
represents a single ANN model that takes 63 density matrix elements as
input features and outputs the probability for each of the six SLOCC
classes. Scheme D (shown in panel (b)) consists of a hierarchical
system of three sequential ANN models to perform SLOCC classification.
ANN Model D1 takes 7 diagonal density matrix elements as input and
predicts whether the state belongs to the FS class. ANN Model D2 then
classifies the states that are non-FS, distinguishing between GME and
BS classes.  Finally, ANN Model D3 further refines the classification
of GME states, separating them into GHZ and W classes. Input layers are
shown in orange, hidden layers in blue, and output layers in green. The
internal structure of each submodel in Scheme D is displayed on the
right.}
\label{GME-SLOCC-figure} 
\end{figure}

The input state is first processed by the FS-NOT model. If the output is
classified as a separable class, the classification task is complete. However,
if the output indicates a non-separable class, it is further analyzed by the
BS-GME model, which classifies it into one of \{BS1, BS2, BS3, GME\}. The
biseparable outputs from this model are considered final, while the GME states
are forwarded to the GHZ-W model. The GHZ-W model then differentiates between
the GHZ and W classes, completing the classification task.

With this hierarchical approach, we achieve an overall test set accuracy of
\(0.940 \pm 0.003\). The accuracies for the GHZ and W classes are \(0.932 \pm
0.004\) and \(0.910 \pm 0.006\), respectively. The performance of this approach
for each class is detailed in Table~\ref{table_averages}. The architecture of
the SLOCC models, along with details of the hidden and output layers of the
submodels, is illustrated in Figure \ref{GME-SLOCC-figure}. The hidden layers
are depicted in blue, while the output layers are shown in green.

\subsection{State Vectors vs Density Matrices as ANN Inputs}
The analysis of the previous section clearly demonstrates that seven features
dominate in determining the presence of GME as well in establishing the
entanglement class for three-qubit pure states.  A natural question that arises
is why did we use  density matrices as inputs to the ANN model and not state
vectors (represented by 15 real parameters), since it is already intuitively
known that there are at most 15 features in the problem!  We hence carried out
the analysis to make a comparison and surprisingly, discovered that that the
ANN was not able to identify the seven important features when given the 15
parameter state vector as an input, and it is in fact advantageous to use
density matrices. 

A three-qubit pure state can be written in the computational
basis with complex coefficients $c_j$ as 
\begin{eqnarray}
|\psi\rangle &=& c_0|000\rangle + c_1|001\rangle +
c_2|010\rangle + c_3|011\rangle + c_4|100\rangle \nonumber \\
&&~~~~+ c_5|101\rangle + c_6|110\rangle +
c_7|111\rangle\nonumber \\
&&{\rm with} \quad \sum_{j=1}^{7}\vert c_j \vert =1. 
\end{eqnarray} 
Thus we have 15
real parameters representing this family of states.
Using the ANOVA-F analysis of these 15 coefficients, we rated the features
according to their importance to the classification task. The top 10
contributors are shown in Table~\ref{table_score_1}.  
As evident from the tabulated values, the difference between
the scores is too low to convincingly 
determine a single minimal set of contributing
features.

\begin{table}[h]
\begin{ruledtabular}
\begin{tabular}{c|cc|cc}
Feature No. & Feature & SLOCC & Feature & GME      \\
\hline
1 & $ \mathbf{Re}(c_7) $ & 2750.58 & $ \mathbf{Re}(c_7) $ & 4388.55\T\\
2 & $ \mathbf{Im}(c_0) $ & 1.45 & $ \mathbf{Re}(c_4) $ & 1.90\\
3 & $ \mathbf{Im}(c_5) $ & 1.42 & $ \mathbf{Re}(c_2) $ & 1.53\\
4 & $ \mathbf{Re}(c_0) $ & 1.41 & $ \mathbf{Im}(c_1) $ & 1.40\\
5 & $ \mathbf{Im}(c_6) $ & 1.27 & $ \mathbf{Re}(c_1) $ & 1.37\\
6 & $ \mathbf{Im}(c_2) $ & 1.25 & $ \mathbf{Im}(c_5) $ & 1.07\\
7 & $ \mathbf{Re}(c_5) $ & 1.24 & $ \mathbf{Im}(c_6) $ & 1.03\\
8 & $ \mathbf{Re}(c_3) $ & 1.03 & $ \mathbf{Re}(c_3) $ & 0.93\\
9 & $ \mathbf{Im}(c_4) $ & 0.95 & $ \mathbf{Im}(c_4) $ & 0.90\\
10 & $ \mathbf{Re}(c_6) $ & 0.84 & $ \mathbf{Im}(c_3) $ & 0.77\\
\end{tabular}
\end{ruledtabular}
\caption{ANOVA F-values (Feature Scores) for the top 10 features
for the SLOCC and GME ANN models.}
\label{table_score_1}
\end{table}

Following the score analysis, we trained the ANN models
using all 15 features, and using the top 9 contributors, and
the top 7 contributors.  The drop in accuracy for the 7
feature model was very significant and hence this model is
not useful. The loss in accuracy when reducing the feature
set to 9 is low for GME detection, but quite significant for
SLOCC classification.  The hierarchical model structure for
SLOCC classification is unable to circumvent this loss.
Therefore, by using the state vector representation for
quantum states, the minimum number of features required for
an ANN model to faithfully classify states remains the same
as the number of unique parameters of the quantum state.
This justifies the use of density matrices in our analysis.
Another reason for using density matrices is that in
real-life situations, for example while working with
experimental data, we may have to work with noisy (mixed)
states, which will anyway require  the states to be
represented in terms of density matrices.
\subsection{Noise Tolerance}
In order to study the noise tolerance of our 
ANN model,  we add
white noise to the states. 
For a three-qubit pure quantum state $\rho$, we
consider the noisy state $\rho^{\prime}$ where a real
positive parameter
$p \leq 1$ determines the amount of noise added to the state:
\begin{equation} 
\rho' = (1-p)\rho + p \mathbb{I} 
\end{equation}
We restrict our analysis to the regime where added noise is a
few percent. The same method is used to add noise 
to the test dataset
and to the training dataset.

We use the ANN trained  on the noise-free training set
to detect and classify entanglement of noise-added states
and we find that the efficiency of the network decreases
very rapidly, even with small amount of added noise.
Table~\ref{nonoise_noise} shows this decline for
the GME as well as the SLOCC classification problem when we
consider a test dataset with 0\%, 1\%, and 2\% noise.

\begin{table}[h]
\begin{tabular}{|c|cc|cc|cc|}
\hline
\multirow{3}{*}{\textbf{Model}} &
\multicolumn{6}{c|}{\textbf{Test Set Noise}}\\ \cline{2-7}
& \multicolumn{2}{c|}{\textbf{0\%}}& 
\multicolumn{2}{c|}{\textbf{1\%}}&
\multicolumn{2}{c|}{\textbf{2\%}}                  \\
\cline{2-7}
                  & \multicolumn{1}{c|}{\textbf{GME}} & 
\textbf{SLOCC} & \multicolumn{1}{c|}{\textbf{GME}} &
\textbf{SLOCC} & \multicolumn{1}{c|}{\textbf{GME}} &
\textbf{SLOCC} \\ \hline
\textbf{63F}      & \multicolumn{1}{c|}{1.00}         & 0.97
& \multicolumn{1}{c|}{0.98} & 0.85    &
\multicolumn{1}{c|}{0.88}         & 0.72           \\ \hline
\textbf{7F}       & \multicolumn{1}{c|}{0.99}        & 0.92 
& \multicolumn{1}{c|}{0.43}   & 0.30           &
\multicolumn{1}{c|}{0.35}         & 0.24           \\ \hline
\textbf{7F Hier}  & \multicolumn{1}{c|}{--}        & 0.93
& \multicolumn{1}{c|}{--}           & 0.28           &
\multicolumn{1}{c|}{--}           & 0.20           \\ \hline
\end{tabular}
\caption{Decline of accuracy of the ANN trained on noiseless
states for detecting entanglement and for the SLOCC
classification of noisy states with different noise values.}
\label{nonoise_noise}
\end{table}

We next consider adding noise to the training dataset and
train the network again with $2\%$ added noise. We then use
this network to detect and classify entanglement of
three-qubit noisy states. The results are summarized in
Table~\ref{noise_noise}. We evaluate the model on a test
dataset with noise values $0\%$, $1\%$ and $2\%$,
respectively.

\begin{table}[h]
\begin{tabular}{|c|cc|cc|cc|}
\hline
\multirow{3}{*}{\textbf{Model}} &
\multicolumn{6}{c|}{\textbf{Test Set Noise}}\\ \cline{2-7}
		  & \multicolumn{2}{c|}{\textbf{0\%}}
& \multicolumn{2}{c|}{\textbf{1\%}}                  &
\multicolumn{2}{c|}{\textbf{2\%}}                  \\
\cline{2-7}
		  & \multicolumn{1}{c|}{\textbf{GME}} &
\textbf{SLOCC} & \multicolumn{1}{c|}{\textbf{GME}} &
\textbf{SLOCC} & \multicolumn{1}{c|}{\textbf{GME}} &
\textbf{SLOCC} \\ \hline
\textbf{63F}      & \multicolumn{1}{c|}{0.99}         & 0.81
& \multicolumn{1}{c|}{0.99}         & 0.91           &
\multicolumn{1}{c|}{1.00}         & 0.98           \\ \hline
\textbf{7F}       & \multicolumn{1}{c|}{0.62}         & 0.26
& \multicolumn{1}{c|}{0.70}         & 0.32           &
\multicolumn{1}{c|}{0.99}         & 0.92           \\ \hline
\textbf{7F Hier}  & \multicolumn{1}{c|}{--}           & 0.30
& \multicolumn{1}{c|}{--}           & 0.38           &
\multicolumn{1}{c|}{--}           & 0.90           \\ \hline
\end{tabular}
\caption{Accuracy of the ANN trained on noisy
states for detecting entanglement and for the SLOCC
classification of noisy states with different noise values.}
\label{noise_noise}
\end{table}
Clearly the network trained with added noise performs much
better when detecting and classifying noisy states. There
also seems to be a correlation between the amount of noise
used for training and the amount of noise present in the
test dataset states. 

Noisy states are mixed states, and mixed quantum states
for a three-qubit system constitute a much larger and
more complex set. It well known that their entanglement
classification is not straightforward. Our aim here was not
to undertake the full fledged problem of entanglement
detection and classification for mixed states, but to
study the noise tolerence of our ANN models 
which were  designed for pure
states.  
\section{Conclusions}
\label{concl}
We demonstrate the effectiveness of an artificial neural
network (ANN) model in accurately classifying a random
three-qubit pure state into one of six SLOCC inequivalent
entanglement classes and detecting the presence of genuine
multipartite entanglement (GME) even with partial data. We
map 35 density matrix elements (7 real and 28 imaginary)
from the pure quantum state of a three-qubit system to 63
ANN input features. The ANN models are trained on a dataset
of $10^6$ numerically generated three-qubit quantum states.
Our results indicate that the ANN model achieves high
accuracies (exceeding 94\%) for both GME detection and SLOCC
classification, using as few as 7 features. Hence, the ANN
model is clearly able to identify a hidden pattern in the
diagonal elements of the three-qubit states. Further, we
evaluate the performance of the ANN models on noisy datasets
by introducing white noise to the test dataset. Our results
demonstrate that the ANN models are robust and well tolerant
towards noise.

Future directions include expanding the model to
higher-dimensional systems and implementing advanced feature
selection techniques and integrating hybrid machine learning
models to improve classification accuracy and efficiency.
Additionally, addressing the robustness of the model against
noise through strategies such as data augmentation and
regularization is essential for practical applications and
work is being carried out in our group on using ANN models
to classify entanglement in mixed states.  Exploring
alternative neural network architectures for specific
quantum platforms can further refine the models and align
them with real-world quantum systems, ultimately paving the
way for real-time classification capabilities.

%

\end{document}